\documentclass{PoS}

\title{Identification of 12 High Mass X-Ray Binaries detected by \textit{INTEGRAL} through NIR photometry and spectroscopy\footnote{Based on observations carried out at the European Southern Observatory under the programs 080.D-0864(A) and 084.D-0535(A), in the visitor mode.}}

\ShortTitle{Identification of new High Mass X-Ray Binaries}

\author{Alexis Coleiro\thanks{This work was supported by the Centre National d'Etudes Spatiales (CNES), based on observations obtained with MINE -- the Multi-wavelength INTEGRAL NEtwork--.}\\
        Laboratoire AIM (UMR-E 9005 CEA/DSM-CNRS-Universit\'e Paris Diderot), Irfu/Service d'Astrophysique, CEA-Saclay\\
        E-mail: \email{alexis.coleiro@cea.fr}}

\author{Sylvain Chaty\\
       Laboratoire AIM (UMR-E 9005 CEA/DSM-CNRS-Universit\'e Paris Diderot), Irfu/Service d'Astrophysique, CEA-Saclay and Institut Universitaire de France, 103 bd Saint Michel, Paris}
        
 \author{Juan A. Zurita Heras\\
Fran\c{c}ois Arago Centre, APC, Universit\'e Paris Diderot, CNRS/IN2P3, CEA/Irfu, Observatoire de Paris, Sorbonne Paris Cit\'e, 10 rue Alice Domon et L\'eonie Duquet, 75205 Paris Cedex 13, France}        
 
 \author{Farid Rahoui\\
   ESO, Karl Schwarzschild-Strasse 2, 85748 Garching bei M\"unchen, Germany}
        
  \author{John Tomsick\\
       Space Science Laboratory, 7 Gauss Way, University of California, Berkeley, CA 94720-7450, USA}

\abstract{Since it started observing the sky, the \textit{INTEGRAL} satellite has revealed new categories of High Mass X-ray binaries in our Galaxy. These observations raise important questions on the formation and evolution of such rare and short-living objects. We present here new observations that enable us to reveal or constrain the nature of 15 new INTEGRAL sources. After previous photometric and spectroscopic campaigns of observations in optical and near infrared, new spectroscopy and photometry were performed at ESO in optical/NIR with EMMI and SOFI on the NTT in 2008 and 2010 on a sample of \textit{INTEGRAL} sources. These observations and precisely the detection of specific features in their spectra allow the identification of such high-energy objects. Our results indicate that: 6 of these objects are Be High-Mass X-ray Binaries (BeHMXBs), 6 are Supergiant High-Mass X-ray Binaries (sgHMXBs) and 3 are still unidentified and need higher resolution data to be completely unveiled.}

\FullConference{"An INTEGRAL view of the high-energy sky (the first 10 years)"
9th INTEGRAL Workshop and celebration of the 10th anniversary of the launch,\\
		October 15-19, 2012\\
		Biblioth\`eque Nationale de France, Paris, France}

\begin{document}

\section{Introduction}
The \textit{INTEGRAL} observatory has been observing the sky for 10 years. By performing a detailed survey of the Galactic plane, it discovered numerous new X-ray binary candidates that need to be accurately identified in order to assess the population of HMXBs and the proportion of supergiant HMXB versus Be HMXBs. Only such a sample will allow us to address the issues related to HMXB formation and evolution. A forthcoming article will present all these results in detail. 
\section{Observations}

The observations that we describe here are based on astrometry, photometry and spectroscopy on 12 INTEGRAL sources.
They were carried out on 2008 March 07-09 and 2010 May 27-29 at the La Silla European Southern Observatory (ESO, Chile), in NIR (1-2.5 $\mu$m) using the SofI instrument installed on the 3.5 m New Technology Telescope (NTT). These observations were performed as the programs ESO ID 080.D-0864(A) and 084.D-0535(A), in the visitor mode. 


\begin{table*}
\scalebox{0.7}{
\begin{tabular}{lllllllll}
\hline
  \hline
  Source & RA & Dec. & Obs. Date & AM & Ks mag. & Derived SpT\\
  \hline

 IGR J10101-5654 & 10:10:11.8 & -56:55:32.0 & 2010-03-29T01:17:11.318 & 1.2  & 10.742 & B0.5Ve  \\
  IGR J11435-6109 & 11:44:00.3 & -61:07:36.5 & 2008-03-09T05:54:12.202 &  1.2  & 11.77  & B0.5Ve \\
 IGR J13020-6359  & 13:01:59.2 & -63:58:05.9  & 2010-03-29T03:16:57.383 & 1.3 & 17.246 & B0.5Ve  \\
 IGR J14331-6112 & 14:33:08.3 & -61:15:39.8 & 2010-03-28T04:53:33.431 & 1.3 &  13.691 & Be \\
 IGR J14488-5942 & 14:48:43.3 &-59:42:16.3 & 2010-03-29T05:56:02.3304&  1.2 & -- & Oe/Be\\
 IGR J16195-4945 & 16:19:32.20 & -49:44:30.7 &  2010-03-30T04:42:57.4456 & 1.5 & -- & ON9.7Iab\\
 IGR J16320-4751 & 16:32:01.9 & -47:52:27.0 & 2010-03-28T06:37:01.7617& 1.2 & --  & BN0.5Ia \\
 IGR J16328-4726 & 16:32:37.91 & -47:23:40.9  & 2010-03-29T07:45:12.428 & 1.1 &  11.309 & O8Iaf or O8Iafpe\\ 
 IGR J16418-4532 & 16:41:50.65 & -45:32:27.3 & 2010-03-28T07:31:53.0447 & 1.1  & -- & BN0.5Ia\\
IGR J17354-3255 & 17:35:27.59  & -32:55:54.4 & 2010-03-30T06:51:03.000 & 1.3 &  10.395 & O8.5Iab(f) or O9Iab \\
IGR J17404-3655 & 17:40:26.8 & -36:55:37.4 & 2010-03-30T08:10:58.684 & 1.1 & 14.370 & sg O/B \\
IGR J17586-2129 & 17:58:34.5 & -21:23:21.6 & 2010-03-30T07:44:31.981 & 1.2 &  9.451 & Be \\ 

\hline
  \label{list_counterparts}
\end{tabular}
}
  \caption{Counterpart positions and summary of the observations carried out in this study.}
  \label{table}
\end{table*}

We performed NIR photometry in Ks band of the sources. For IGR J11435-6109, J and H bands photometry was also carried out. The photometric observations were obtained by repeating a set of images for each filter with 9 different 30" offset positions including the targets, following the standard jitter procedure that enable us to cleanly subtract the sky emission in NIR. Each individual frame has an integration time of 10 s, giving a total exposure time of 90 s in each energy band. Moreover, three photometric standard stars chosen in the Person's catalogue were observed in the 3 bands with a total integration time of 10 s for each one in each band. We also carried out NIR spectroscopy with SofI. In 2008, 8 spectra were taken using the low-resolution red grism (R = 980 for 1.53 $\mu$m < $\lambda$ < 2.52 $\mu$m), half of them with the 1".0 slit on the source and the other half with an offset, in order to subtract the NIR sky emission. Each individual spectrum has an exposure time of 60 s, giving a total integration time of 480 s. Furthermore, four telluric standards were observed during the observing night with the same instrument setup and a total integration time of 8 s. Second observing run used the medium resolution grism (R = 2200 for 2.00 $\mu$m < $\lambda$ < 2.30 $\mu$m) with the H and Ks filters. Each individual spectrum has an exposure time of 60 s. 4 telluric standards were observed.


\section{Data reduction}

We used the IRAF (Image Reduction and Analysis Facility) suite to perform data reduction, carrying out standard procedures of NIR image reduction, including flat-fielding and NIR sky subtraction.We performed accurate astrometry on each entire SofI 4'.92 $\times$ 4'.92 field, using all stars from the 2MASS catalogue present in this field. We carried out aperture photometry on photometric standard stars in order to compute the Zero-point value for each energy band. Then, we performed PSF-fitting photometry on the crowded fields following the standard utilization of the IRAF \textit{noao.digiphot.daophot} package. Using an aperture correction for each filter and each object, we evaluated the apparent magnitude of the targets in the three NIR filters. The results are given in Table \ref{table}. Then, we analyzed the NIR spectra using standard IRAF tasks, correcting for flat field, removing the crosstalk, correcting the geometrical distortion, combining the images and finally extracting the spectra and performing wavelength calibration using the IRAF \textit{noao.twodspec} package. Wavelength calibration was done with Xenon lamp presenting a good distribution of lines and sufficient to calibrate data taken with the low resolution grisms of SofI instrument. The targets spectra were then corrected for the telluric lines using the standard stars observed with the same configuration. 

\section{Results}

All the sources studied in this paper were discovered with the IBIS/ISGRI detector onboard the INTEGRAL observatory. We present in the following our results on each source, for which we followed the same strategy. We first observed the field in NIR, we performed accurate astrometry, and we derived the photometry of the counterpart candidate. We then analyzed the NIR spectrum. 

\subsection{IGR J10101-5654}

\cite{Masetti2006} identified a candidate optical counterpart that was later localized via Chandra observations (\cite{Tomsick2008}). The optical spectrum is typical of an HMXB star, showing strong narrow H$_\alpha$ emission superimposed on a reddened continuum.  They suggest this HMXB to host a secondary star of intermediate luminosity class (early giant). Our NIR H band spectrum exhibits the presence of the Brackett series from Br(20-4) line at 1.5198 $\mu$m to Br(10-4) transition at 1.7377 $\mu$m. Ks band spectrum clearly shows the presence of HeI at at 2.0586 $\mu$m and Br(7-4) at 2.1659 $\mu$m. MgII, HeII, NaI, FeI and MgI emission lines are also present in this spectrum. The NIR spectra are typical of a Be companion star and the intensity ratio of the Br(7-4) and HeI at 2.0586 $\mu$m lines suggests a B0.5Ve type (\cite{Hanson1996}).

\subsection{IGR J11435-6109}

The optical and infrared counterpart was discovered by \cite{Tomsick2007} and confirmed by \cite{Negueruela2007_1} as the 2MASS J11440030-6107364. The spectral type was constrained by \cite{Masetti2009} who suggested a B2III or B0V counterpart with an extinction A$_\mathrm{V}$ of 5.7. Our NIR H band spectrum exhibits the presence of the Brackett series from Br(21-4) line at 1.5137 $\mu$m to Br(10-4) transition at 1.7364 $\mu$m. Ks band spectrum clearly shows the presence of HeI at at 2.0590 $\mu$m and Br(7-4) at 2.1664 $\mu$m. MgII and HeII emission lines are also present in this spectrum. These NIR spectra are pretty similar to the ones of IGR J10101-5654, typical of a Be companion star and the intensity ratio of the Br(7-4) and HeI (at 2.0590 $\mu$m) lines suggests a B0.5Ve type (according to \cite{Hanson1996}).

\subsection{IGR J13020-6359}

\cite{Chernyakova2005} identified the likely 2MASS infrared counterpart. Ks band spectrum of the source clearly shows the presence of HeI at at 2.0594 $\mu$m and Br(7-4) at 2.1663 $\mu$m. Even if other lines are not detected, the NIR spectrum is typical of a Be companion star and the intensity ratio of the Br(7-4) and HeI (at 2.0594 $\mu$m) lines suggests a B0.5Ve type (\cite{Hanson1996}), similar of IGR J10101-5654 and IGR J11435-6109.

\subsection{IGR J14331-6112}

\begin{figure}[th]
\begin{center}
\includegraphics[scale=0.25]{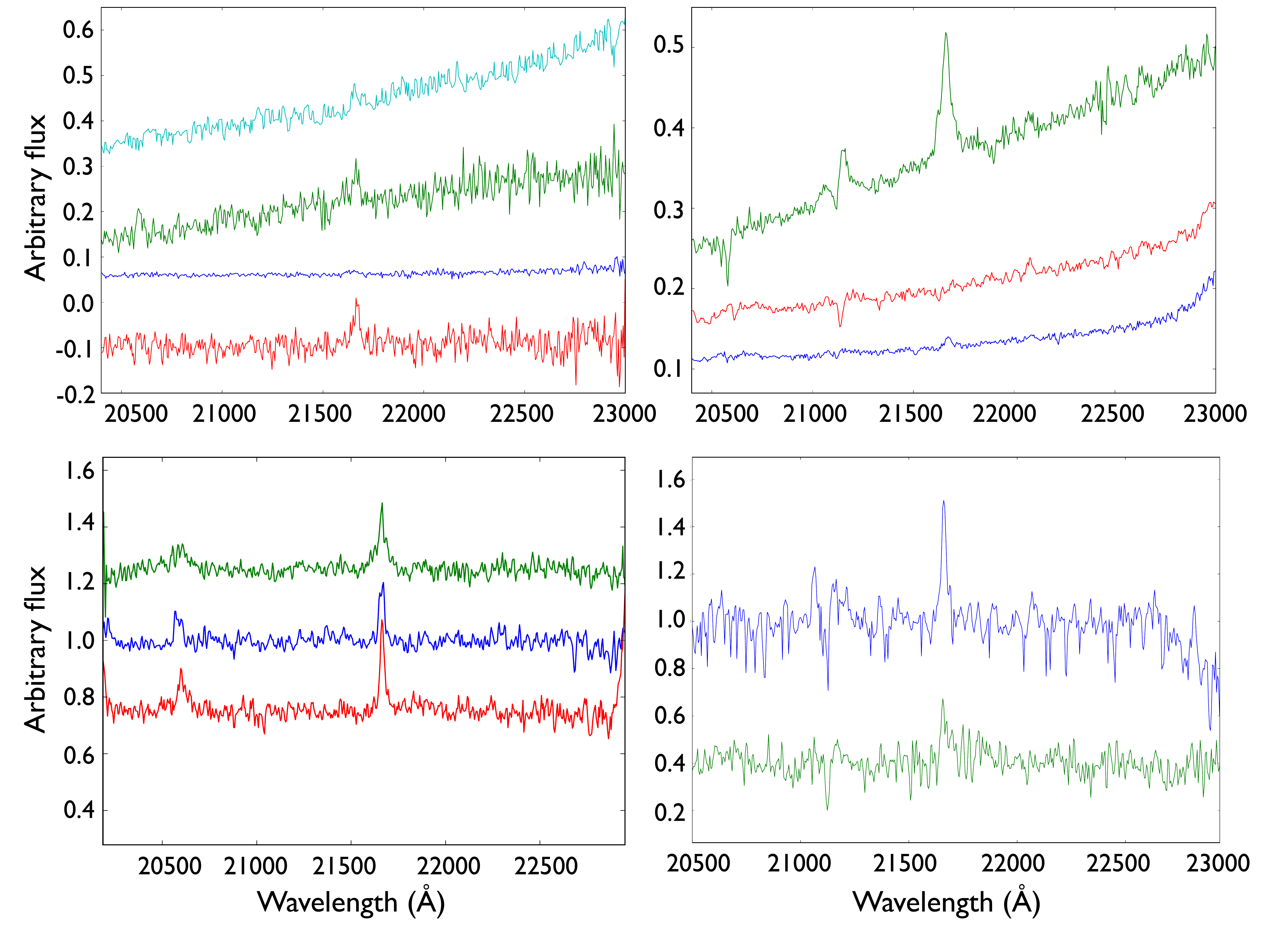}
\caption{Ks band spectra of the 12 sources. Left upper panel (BeHMXBs), cyan: IGR J17586-2129; green: IGR J14488-5942; blue: IGR J14331-6112; red: IGR J17404-3655. Right upper panel, green: IGR J16328-4726; red: IGR J17354-3255; blue: IGR J16195-4945. Left lower (B0.5Ve), green: IGR J11435-6109; blue: IGR J10101-5654; red: IGR J13020-6359. Right lower (BN0.5Ia), blue: IGR J16320-4751; green: IGR J16418-4532. }
\label{spectra}
\end{center}
\end{figure}

Optical spectroscopy carried out by \cite{Masetti2008} indicated an HMXB hosting a companion star of BIII or a BV spectral type. Ks band spectrum of the source is shown on Figure \ref{spectra}, left upper panel. Even if the signal to noise ratio of this spectrum is low, it shows the presence of a pretty large emission line at 2.1657 $\mu$m that could be the Br(7-4) line. Then, we cannot conclude definitively on the spectral type of the companion star of IGR J14331-6112 but according to the Br(7-4) emission line, it could be a Be star.

\subsection{IGR J14488-5942}

2MASS 14484322-5942137 seems to be the near infrared counterpart with J=15.46 mag, H=13.53 mag and Ks=12.43 mag (see \cite{Landi2009}). \cite{Corbet2010} analyzed the 15-100 keV light curve of this source and detected a highly significant modulation at a period near 49 days. This long term variation is interpreted as the orbital period of an HMXB (possibly a BeHMXB). Ks band spectrum of the source only shows potential emission line of HeI at 2.0590 $\mu$m and Br(7-4) at 2.1665 $\mu$m. Again, we cannot conclude definitely on the spectral type of the companion star but it could be an Oe/Be star.

\subsection{IGR J16195-4945}

The refined position determined by \cite{Tomsick2006}) enables them to find the NIR and MIR counterparts in the 2MASS (2MASS J16193220-4944305) and in the GLIMPSE (G333.5571+00.3390) catalogues. According to \cite{Rahoui2008}, the stellar component is consistent with an O/B supergiant. On the Ks band spectrum we only observe a pretty large emission line at 2.1672 $\mu$m that can be the Br(7-4) emission line. Two suspected absorption line at 2.0577 $\mu$m and at 2.1120 $\mu$m could be HeI absorption lines. In that case, according to (\cite{Hanson2005}), it could be an ON9.7Iab companion star.

\subsection{IGR J16320-4751}

NIR observations of the most likely counterpart conducted by \cite{Chaty2008} suggest a luminous supergiant OB companion star. SED fitting computed in \cite{Rahoui2008} and observations of \cite{Chaty2008} finally state that this source belongs to the very obscured supergiant HMXB class, hosting a neutron star. In our Ks band spectrum, emission line of Br(7-4) is clearly detected at 2.1668 $\mu$m. HeI emission line at 2.0586 $\mu$m is weak but quite broad whereas another HeI absorption line is detected at 2.1127 $\mu$m. These lines lead to a classification of the companion star as a BN0.5Ia according to \cite{Hanson2005}.

\subsection{IGR J16328-4726}

According to \cite{Fiocchi2010}, the source 2MASS J16323791-4723409 is the most likely candidate IR counterpart. \cite{Fiocchi2013} claim that NIR colors are compatible with both OB supergiant and OB main sequence companion star. Our Ks band spectrum shows that both HeI emission line at 2.1152 $\mu$m and Br(7-4) emission line at 2.1661 are clearly detected. Moreover, HeI at 2.0579 is detected in absorption whereas HeI in emission is also detected at 2.1152 $\mu$m together with NIII/CIII emission around 2.116 $\mu$m. According to \cite{Hanson2005} it can be a O8Iaf or more probably a O8Iafpe taking into account the HeI / Br(7-4) line ratio. 

\subsection{IGR J16418-4532}

Near Infrared (NIR) observations conducted by \cite{Chaty2008} proposed 4 NIR  candidate counterparts and the brightest one, 2MASS J16415078-4532253, is favoured to be the counterpart. SED fitting carried out by \cite{Chaty2008} and \cite{Rahoui2008} suggest an OB spectral type companion star with enshrouding material that marginally contributes to its MIR emission. NIR Ks band spectrum is presented in Figure \ref{spectra}, left lower panel. It shows a wide emission line at 2.1672 $\mu$m that corresponds to the Br(7-4) line and a weak emission line of HeI at 2.0580 $\mu$m. Moreover, we observe an absorption line of HeI at 2.1124 $\mu$m. These detections point towards a BN0.5Ia spectral type\footnote{We point out that X-shooter observations (see \cite{Goldoni2013}) show Br(7-4) line in absorption and then lead to a O9.5I spectral classification.}.

\subsection{IGR J17354-3255}

\cite{Vercellone2009} and \cite{Tomsick2009_1} observed the counterpart as 2MASS J17352760-3255544. \textit{XMM-Newton} observations carried out by \cite{Bozzo2012} suggest an SFXT behaviour. NIR Ks band spectrum, presented in Figure \ref{spectra}, right upper panel, shows a clear HeI absorption at 2.1134 $\mu$m. A weak absorption is detected around 2.06 $\mu$m that could be an HeI absorption. Then, this spectrum points towards a O8.5Iab(f) (see \cite{Hanson1996}) or a O9Iab (see \cite{Hanson2005}) spectral type but the low signal to noise ratio prevents any better classification.

 \subsection{IGR J17404-3655}
 According to \cite{Landi2008}, the optical counterpart of this source is USNO-A2.0 0525\_28851523. \textit{Chandra} observations, conducted by \cite{Tomsick2009}, claim that this source is probably an HMXB. Only the Br(7-4) line is detected in emission in our NIR Ks band spectrum. According to the high equivalent width value of the Br$\gamma$ emission line, it could be a supergiant star. This would confirm the HMXB type of the binary system ruling out the LMXB classification.

\subsection{IGR J17586-2129}
\cite{Tomsick2009} identified the infrared counterpart of this source as 2MASS J17583455-2123215. It is suggested to be candidate for an absorbed HMXB. Again, only the Br(7-4) line is detected in emission in our NIR Ks band spectrum. The companion star could be a Be star but the low signal to noise ratio prevents any better classification.

%

%

%

\section{Conclusion}
We have presented the results of our last NIR photometry and spectroscopy observations of a sample of unidentified IGR sources. We determined or improved the spectral classification of the companion star in twelve cases. Some targets need higher resolution spectroscopy before firmly confirming the identification. These results enable us to improve the HMXB population studies.  

\bibliographystyle{plain}
\bibliography{biblio1}

\end{document}